\documentclass[letterpaper, 10 pt, conference]{ieeeconf} 
\IEEEoverridecommandlockouts
\usepackage{cite}
\usepackage{hyperref}
\hypersetup{
	colorlinks=true,
	linkcolor=blue,
	citecolor=black,
}
\usepackage{enumerate}
\usepackage{amsmath,amssymb,amsfonts}
\usepackage{algorithm,algorithmicx,algpseudocode}
\usepackage{graphicx}
\usepackage{subcaption}
\usepackage{xcolor}

\newcommand{\argmax}{\arg\!\max}

\DeclareMathOperator*{\diag}{\textit{diag}}
\DeclareMathOperator*{\trace}{\textit{tr}}

\newtheorem{program}{\bf Program}[section]
\newtheorem{theorem}{\bf Theorem}[section]

\newtheorem{definition}{\bf Definition}[section]
\newtheorem{remark}{\bf Remark}[section]%{\hfill$\Box$}

\newenvironment{pf}{{\it  Proof:}}{\hfill$\Box$}
\newcommand{\nnum}{\nonumber}
\newcommand{\vect}[1]{\mathbf{#1}}

\begin{document}
%%%%%%%%%%%%%%%%%%%%%%%%%%%%%%%%%%%%%%%%%%%%%%%
%%%%%%%%%%%% Title & Authors %%%%%%%%%%%%%%%%%%
%%%%%%%%%%%%%%%%%%%%%%%%%%%%%%%%%%%%%%%%%%%%%%%
\title{\LARGE \bf
A Safety Constrained Control Framework for UAVs in GPS Denied Environment
}
%%%%%%%%%%%%%%%%%%%%%%%%%%%%%%%%%%%%%%%%%%%%%%%
\author{Wenbin Wan$^{\dagger}$, Hunmin Kim$^{\dagger}$, Naira Hovakimyan$^{\dagger}$, Lui Sha$^{*}$ and Petros G. Voulgaris$^{\ddagger}$% <-this % stops a space
\thanks{This work has been supported by the National Science Foundation (ECCS-1739732 and CMMI-1663460).}
\thanks{$^{\dagger}$Wenbin Wan, Hunmin Kim, and Naira Hovakimyan are with the Department of Mechanical Science and Engineering, University of Illinois at Urbana-Champaign, USA.
{\tt\small  \{wenbinw2, hunmin, nhovakim\}@illinois.edu }}%
\thanks{$^{*}$Lui Sha is with the Department of Computer Science, University of Illinois at Urbana-Champaign, USA.
{\tt\small  lrs@illinois.edu }}%
\thanks{$^{\ddagger}$Petros G. Voulgaris is with the Department of Mechanical Engineering, University of Nevada, Reno, USA.
{\tt\small  pvoulgaris@unr.edu }}%
}
%%%%%%%%%%%%%%%%%%%%%%%%%%%%%%%%%%%%%%%%%%%%%%%
%%%%%%%%%%%%%%%%%%%%%%%%%%%%%%%%%%%%%%%%%%%%%%%
\maketitle

\begin{abstract}
Unmanned aerial vehicles (UAVs) suffer from sensor drifts in GPS denied environments, which can lead to potentially dangerous situations.
To avoid intolerable sensor drifts in the presence of GPS spoofing attacks, we propose a safety constrained control framework summarized in Fig.~\ref{fig:demo} that adapts the UAV at a path re-planning level to support resilient state estimation against GPS spoofing attacks.
The attack detector is used to detect GPS spoofing attacks based on the resilient state estimation and provides a switching criterion between the robust control mode and emergency control mode.
To quantify the safety margin, we introduce the escape time which is defined as a safe time under which the state estimation error remains within a tolerable error with designated confidence.
An attacker location tracker (ALT) is developed to track the location of the attacker and estimate the output power of the spoofing device by the unscented Kalman filter (UKF) with sliding window outputs.
Using the estimates from ALT, an escape controller (ESC) is designed based on the constrained model predictive controller (MPC) such that the UAV escapes  from the effective range of the spoofing device within the escape time.
\end{abstract}

\section{Introduction}
UAVs have been used across the world for commercial, civilian, as well as educational applications over the decades.
The mechanical simplicity and agile maneuverability appeal to many applications, such as cargo transportation~\cite{maza2009multi}, aerial photography~\cite{hardin2005unmanned}, and agricultural farming~\cite{candiago2015evaluating}.
The most widely used sensor for UAVs is the global positioning system (GPS), which offers accurate and reliable state measurements.
However, GPS receivers are vulnerable to various types of attacks, such as blocking, jamming, and spoofing~\cite{warner2003gps}. 
The Vulnerability Assessment Team at Los Alamos National Laboratory has demonstrated that the civilian GPS spoofing attacks can be easily implemented by using GPS simulator~\cite{warner2002simple}.
Furthermore, GPS is more vulnerable when its signal strength is weak.
Due to various applications of UAVs, the operating environment becomes diverse as well, where GPS signals are weak or even denied due to other structures such as skyscrapers, elevated highways, bridges, and mountains.

\begin{figure}[!t]
\centering
 \includegraphics[width=0.35\textwidth]{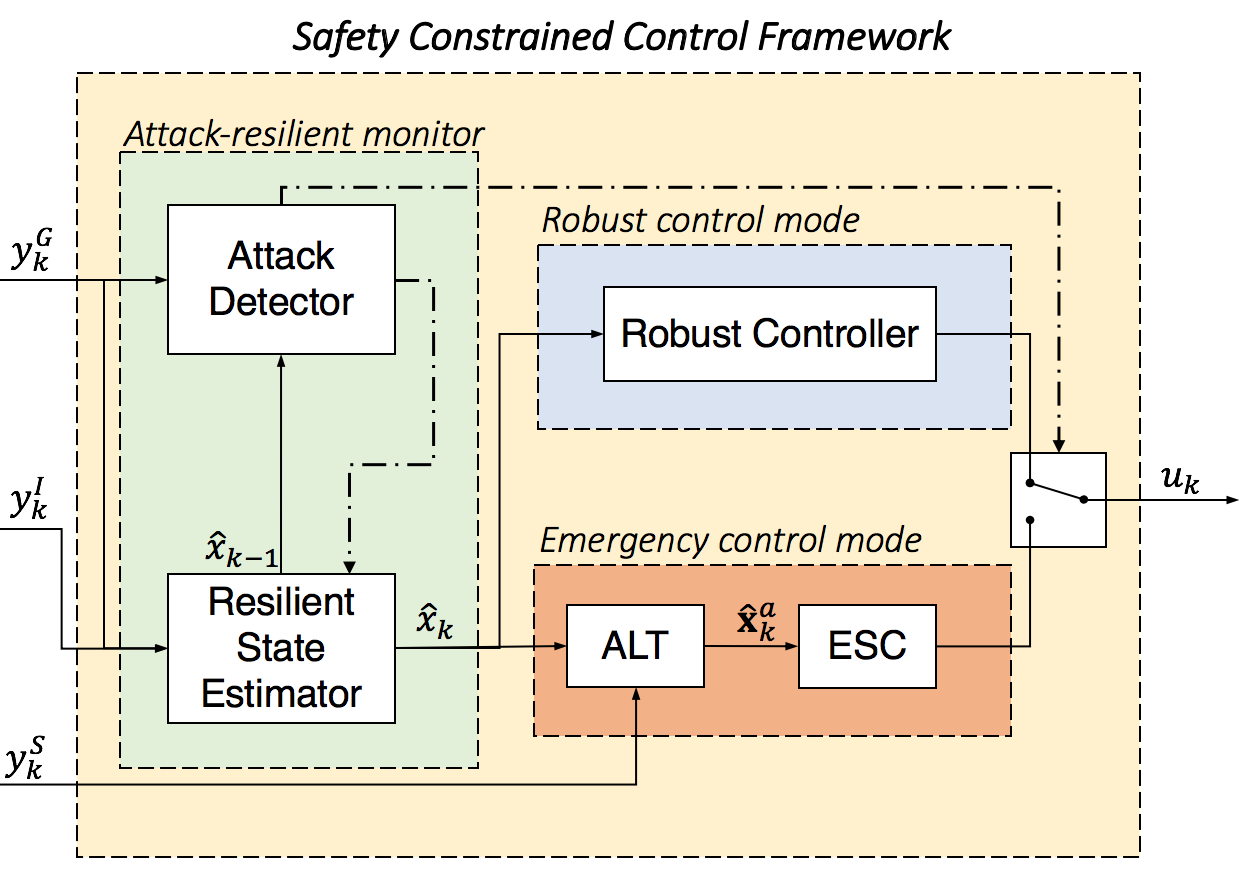}
\caption{\it A safety constrained control framework consisting of an attack detector, a resilient state estimator, a robust controller, an ALT and an ESC.} \label{fig:demo} \vspace{-.7cm}
\end{figure}

\emph{Literature review.}
One of the GPS spoofing attack detection techniques is to analyze raw antenna signals or utilize multi-antenna receiver systems.
The GPS spoofing attack can be detected by checking whether the default radiation pattern is changed in~\cite{mcmilin2014single}.
A multi-antenna receiver system was used to detect GPS spoofing attacks by monitoring the angle-of-arrival of the spoofing attempts in~\cite{montgomery2011receiver}.
As an extension of this work, the GPS spoofing mitigation has also been investigated where an array of antennas is utilized to obtain genuine GPS signals by spatial filtering~\cite{magiera2015detection,chen2012study,chen2013validation}.
However, those solutions require modifications of the hardware or the low-level computing modules and assume that an attacker can only use single-antenna spoofing systems.
Furthermore, the attacker can spoof the GPS receivers without being detected if multi-antenna spoofing devices are available~\cite{jansen2017advancing}.

In CPS security literature, GPS spoofing attacks have been described as a malicious signal injection to the genuine sensor output~\cite{mo2010false}.
Attack detection against malicious signal injection has been widely studied over the last few years.
The attack detection problem has been formulated as an $\ell_0$/$\ell_\infty$ optimization problem, which is NP-hard in~\cite{fawzi2014secure,pajic2014robustness}.
The fundamental limitations of structural detectability, as well as graph-theoretical detectability for linear time invariant systems have been studied in~\cite{pasqualetti2013attack}, where distributed attack detection has also been studied.
The attack detection problem has been formulated as an attack-resilient estimation problem of constrained state and unknown input in~\cite{wan2019attack}.
A switching mode resilient detection and estimation framework for GPS spoofing attacks has been studied in~\cite{yoon2019towards}.
Attack detection using multiple GPS signals by checking cross-correlation was introduced in~\cite{psiaki2013gps}.
In~\cite{5718158}, the maximum deviations of the state were identified due to the sensor attacks while remaining stealthy due to the detection.
A secure control framework for networked control systems was designed in~\cite{teixeira2015secure} to analyze the resource-limited adversaries.
We notice that existing emergency control architectures focus on switching control from a high-performance controller to a robust high-assurance controller in the presence of attacks~\cite{wang2018rsimplex}.
These architectures can efficiently handle a class of attacks, but cannot address the fundamental problem due to limited sensor availability in the presence of cyber-attacks.

\emph{Contribution.} 
The current paper addresses safety problems induced by limited sensor availability due to GPS spoofing attacks.
We formulate the sensor drift problem as an increasing variance of state estimation to quantify the sensor drift and introduce escape time under which the state estimation error remains within a tolerable error with high confidence.
We develop a novel safety constrained control framework that adapts the UAV at a path re-planning level to support resilient state estimation against GPS spoofing attacks.
In the presence of the GPS spoofing attack, the attacker location tracker (ALT) tracks the attacker's location and estimates the output power of the spoofing device by UKF with sliding window outputs.
The estimates are then used in the escape controller (ESC) that drives the UAVs away from the effective range of the spoofing device within the escape time to avoid intolerable sensor drift.

\section{Preliminaries} \label{sec:pre}
\subsection{Notation}
We use the subscript $k$ of $x_k$ to denote the time index;
${\mathbb R}^n_+$ denotes the set of positive elements in the n-dimensional Euclidean space;
${\mathbb R}^{n \times m}$ denotes the set of all $n \times m$ real matrices;
$A^\top$ $A^{-1}$ and $\trace(A)$ denote the transpose, inverse and trace of matrix $A$, respectively;
$I$ denotes the identity matrix with an appropriate dimension;
$\|\cdot\|$ denotes the standard Euclidean norm;% for vector or an induced matrix norm;
$\times$ is used to denote matrix multiplication when the multiplied terms are in different lines;
${\mathbb E}[\,\cdot\,]$ denotes the expectation operator;
${\mathbb P}[\,\cdot\,]$ denotes the probability operator.
For a matrix $S$, $S > 0$ and $S \geq 0$ indicate that $S$ is positive definite and positive semi-definite, respectively.
 
\subsection{System Model}
Consider the discrete-time dynamic system model:
\begin{subequations} \label{eq1 sys}
\begin{align} 
    x_{k} &= A x_{k-1} + B u_{k-1} + w_{k-1} \\
    y_k^G &= C^G x_k  +  d_k + v_k^G \label{eq1 b}\\
    y_k^I &= C^I (x_k - x_{k-1}) + v_k^I \label{eq1 c}\\
    y_k^S &=
            \left\lbrace
            \begin{array}{cc}
            C^S \frac{\eta_k}{d(x^a_k, x_k)^2}+ v_k^S, &\textit{under the attack} \vspace{+.2 cm}\\
            \eta^{S} + v_k^{S}, & \textit{otherwise ,}
            \end{array}
            \right.
                \label{eq1 d} 
            \end{align}
\end{subequations}
where $x_k \in {\mathbb R}^n$ is the state, and $A$, $B$, $C^G$, $C^I$ and $C^S$ are proper sized matrices.
There are three types of outputs available.
Output $y_k^G \in {\mathbb R}^{m_G}$ is the GPS measurement which may be corrupted by unknown GPS spoofing signal $d_k \in {\mathbb R}^{m_G}$.
The signal $d_k$ is injected by the attacker when the UAV is in the effective range of the spoofing device.
Output $y_k^I \in {\mathbb R}^{m_I}$ is the IMU measurement which returns a noisy measurement of the state difference.
Output $y_k^S \in {\mathbb R}^{m_S}$ represents the GPS signal strength.

The defender is unaware of $x_k^a$ and $\eta_k$, where $x_k^a \in {\mathbb R}^n$ is the unknown attacker location, and $\eta_k \in {\mathbb R}^{m_S}$ is the nominal power of the spoofing device. 
If GPS is under the attack, $y_k^S$ is an inverse function of the distance between the attacker and UAV.
The function $d(a,b)$ measures the distance between $a$ and $b$.
If the UAV receives genuine GPS signals, this output represents the genuine GPS signal strength $\eta^{S}$. 
We assume that the attacker can inject any signal $d_k$ into $y_k^G$.

The noise signals $w_k$, $v_k^G$, $v_k^I$, and $v_k^S$ are assumed to be independent and identically distributed Gaussian random variables with zero means and covariances
${\mathbb E[w_k w_k^\top]=\Sigma}_w \geq 0$,
${\mathbb E[v_k^G (v_k^G)^\top]=\Sigma}_G >0$,
${\mathbb E[v_k^I (v_k^I)^\top]=\Sigma}_I>0$,
and ${\mathbb E[v_k^S (v_k^S)^\top]=\Sigma}_S>0$, respectively.

\begin{remark}
The sensor measurement $y_k^I$ represents any  {\em relative} sensor measurement, such as velocity measurement by a camera. In this paper, we use IMU for the illustration.
\end{remark}

\begin{remark}
The signal strength output $y_k^S$ in~\eqref{eq1 d} is derived by the GPS signal attenuation due to free-space path loss.
Friis transmission equation is given by:
$P_{r}=P_{t}G_{t} G_{r} \frac{\lambda^{2}}{(4 \pi r)^{2}}$,
where $P_{t}$ and $P_{r}$ are the transmit power and the receive power; $G_{t}$ and $G_{r}$ are the transmit and receive antenna gains; $r$ is the distance between two antennas; $\lambda$ is the wavelength~\cite{1697062}.
We write $G_r (\frac{\lambda}{4\pi})^2$ as the output matrix $C_S$ ; $G_tP_t$ as the nominal power of the spoofing device $\eta_k$; and $r$ as the distance $d(x^a_k, x_k)$.
\end{remark}

\subsection{Problem Statement}\label{sec:pro}
Given the system~\eqref{eq1 sys} with sensor measurements~\eqref{eq1 b}-\eqref{eq1 d}, the defender aims to detect the GPS spoofing attack, achieve resilient state estimation when considering the limited sensor availability,
and complete the global mission securely.

\section{Safety Constrained Control Framework} \label{sec:framework}
To address the problem described in Section~\ref{sec:pro}, we propose a safety constrained control framework in Fig.~\ref{fig:demo},
which consists of an attack detector, a resilient state estimator, a robust controller, an attacker location tracker (ALT), and an escape controller (ESC). The proposed safety constrained control framework drives the UAV to the outside of the effective range of the spoofing device.
The following explains each module in the proposed framework as shown in Fig.~\ref{fig:demo}.

\noindent \emph{Robust Control Mode.}
The robust controller is a complex controller that operates the UAV to the destination in the presence of noise, but without the presence of attacks.
Any robust control technique can be implemented to this module.

\noindent \emph{Emergency Control Mode.}
ALT is designed for tracking the location of the attacker and estimating the output power of the spoofing device by applying UKF with sliding window outputs.
ESC is an MPC-based controller that drives the UAV out of the effective range of the spoofing device based on the estimation of the attacker location obtained by ALT.

\noindent \emph{Attack-resilient Monitor \& Decision Logic.}
The resilient state estimator is developed based on the Kalman-filter like state estimator.
The attack detector is designed by the $\chi^2$-based anomaly detection algorithm.
Based on the previous estimation from the resilient state estimator, the Boolean output (dotted-dashed line in Fig.~\ref{fig:demo}) of the attack detector determines
$(i)$ whether the GPS measurement should be used for the state estimation; and $(ii)$ the switching rule between two control modes: the robust control mode and the emergency control mode.

ALT and ESC adapt the UAV at a path re-planning level for safe operation.
In what follows, each subsection describes the details of the corresponding component. 

\subsection{Resilient State Estimator}\label{sec:estimation}
The defender implements an estimator and $\chi^2$ detector to estimate the state and detect the GPS spoofing attack.
The following Kalman-filter like state estimator is used to estimate the current state:
\begin{align}
\hat{x}_{k} &= A \hat{x}_{k-1}+ B u_{k-1}
+ K_k^G (y_k^G - C^G( A\hat{x}_{k-1} + B u_{k-1})) \nonumber  \\
&+ K_k^I (y_k^I  -C^I( A \hat{x}_{k-1} + B u_{k-1}-\hat{x}_{k-1})) \label{eq x estim}\\
P_k&= (A-K_kCA+K_kDC)P_{k-1}
 (A-K_kCA+K_kDC)^\top \nonumber  \\
&+(I-K_kC)\Sigma_w (I-K_kC)^\top+K_k \Sigma_y K_k^\top, \label{eq P_update}
\end{align}
where $\hat{x}_k$ is the state estimate and $P_k$ is the state estimation error covariance at time $k$, and
$K_k :=\left[ \begin{array}{cc} K_k^G& K_k^I \end{array} \right]$,
$C:= \left[ \begin{array}{c}  C^G\\ C^I\\ \end{array} \right]$,
$\Sigma_y := \left[ \begin{array}{cc} \Sigma_G& 0\\ 0&\Sigma_I\\ \end{array} \right]$ and
$D := \left[ \begin{array}{cc} 0&0\\ 0&I\\ \end{array} \right]$.
The optimal gain $K_k$, given by
\begin{align}\label{eq:K_law}
&K_k=(AP_{k-1}(CA-DC)^\top+\Sigma_wC^\top)\\
&\times\left((CA-DC)P_{k-1}(CA-DC)^\top\right.+\left.C\Sigma_wC^\top+\Sigma_y\right)^{-1},\nonumber
\end{align}
is the solution of the optimization problem $\min_{K_k} \trace{(P_k)}$.

In~\cite{yoon2019towards}, it has been shown that the covariance in~\eqref{eq P_update} is bounded when the GPS signal is available. If the GPS is denied, and only the relative sensor $y_k^I$ is available, the covariance is strictly increasing and unbounded in time. That is, the sensor drift problem can be formulated as  instability of the covariance matrix.

\subsection{Attack Detector}
We conduct the $\chi^2$ test to detect the GPS spoofing attacks:
\begin{equation}
     H_0: d_k=0; \quad H_1: d_k \neq 0, \label{e000.1}
\end{equation}
using CUSUM (CUmulative SUM) algorithm, which is widely used in attack detection research~\cite{page1954continuous,barnard1959control,lai1995sequential}.

Since $d_k = y_k^G - C^G x_k - v_k^G$, given the previous state estimate $\hat{x}_{k-1}$, we estimate the attack vector by comparing the sensor output and the output prediction:
\begin{align}
    \hat{d}_k     &=y_k^G - C^G ( A\hat{x}_{k-1} + B u_{k-1}). \label{est d_k}
\end{align}
Note that the current estimate $\hat{x}_k$ should not be used for the prediction, because it is correlated with the current output; i.e., ${\mathbb E}[\hat{x}_k (y_k^G)^\top] \neq 0$.
Due to the Gaussian noises $w_k$ and $v_k$ injected to the linear system in~\eqref{eq1 sys}, the states follow Gaussian distribution since any finite linear combination of Gaussian distributions is also Gaussian.
Similarly, $\hat{d}_k$ is Gaussian as well, and thus the use of $\chi^2$ test~\eqref{e000.1} is justified.
In particular, the $\chi^2$ test compares the normalized attack vector estimate $\hat{d}_k^\top (P_{k}^d)^{-1}\hat{d}_k$ with $\chi^2_{df}(\alpha)$:
\begin{equation}
\begin{aligned}
&\text { Accept $H_0$, if } \hat{d}_k^\top (P_{k}^d)^{-1}\hat{d}_k \leq \chi^2_{df}(\alpha)\\
&\text { Accept $H_1$, if } \hat{d}_k^\top (P_{k}^d)^{-1}\hat{d}_k > \chi^2_{df}(\alpha),
\end{aligned}\label{e003}    
\end{equation}
where $P_{k}^d := {\mathbb E}[(d_k-\hat{d}_k)(d_k-\hat{d}_k)^\top]=C^G(A P_{k-1}A^\top+\Sigma_w)(C^G)^\top+\Sigma_G$, and $\chi_{df}^2(\alpha)$ is the threshold found in the Chi-square table. In $\chi_{df}^2(\alpha)$, $df$ denotes the degree of freedom, and $\alpha$ denotes the statistical significance level.

To reduce the effect of noise, we use the test~\eqref{e003} in a cumulative form.
The proposed $\chi^2$ CUSUM detector is characterized by the detector state $S_k\in\mathbb{R}_{+}$:
\begin{align}
    S_{k}=\delta S_{k-1}+(\hat{d}_k)^\top (P_{k}^d)^{-1}\hat{d}_k, \quad S_0=0,
    \label{e003.1}
\end{align}
where $0<\delta<1$ is the pre-determined forgetting factor. 
At each time $k$, the CUSUM detector~\eqref{e003.1} is used to update the detector state $S_k$ and detect the attack.

The attack detector will $i)$ update the estimated state $\hat{x}_k$ and the error covariance $P_k$ in~\eqref{eq P_update} with $K_k^G=0$ and $ii)$ switch the control mode to emergency control mode,
if
\begin{align}
     S_k>\sum_{i=0}^{\infty}\delta^i\chi^2_{df}(\alpha)=\frac{\chi^2_{df}(\alpha)}{1-\delta}.\label{e003.2}
\end{align}
If $S_k<\frac{\chi^2_{df}(\alpha)}{1-\delta}$, then it returns to the robust control mode. 

\begin{figure}[htpb]
\centering
 \includegraphics[width=0.36\textwidth]{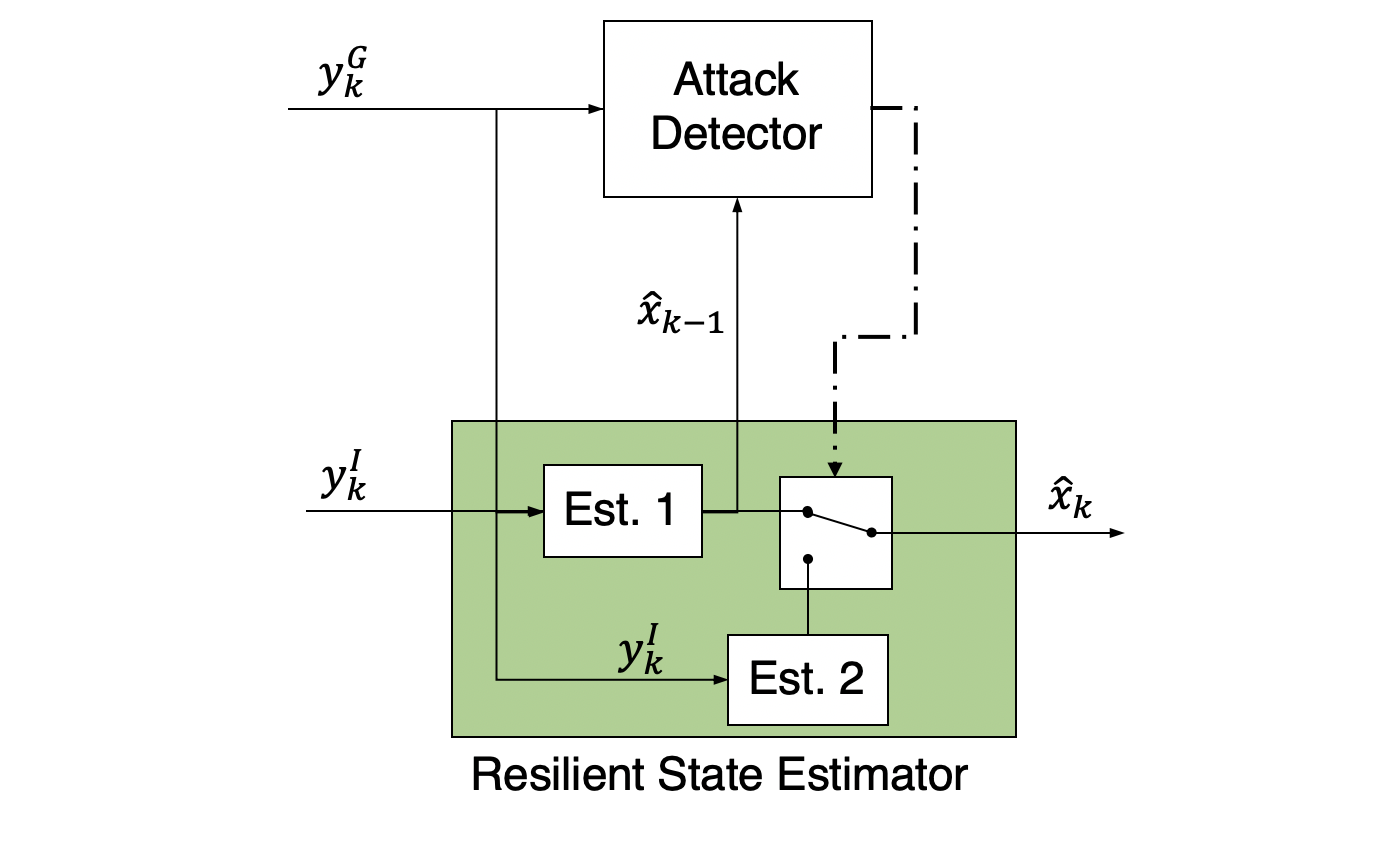}\vspace{-0.3cm}
\caption{\it Resilient state estimator. GPS and IMU measurements are used in the estimator one (Est. 1). Estimator two (Est. 2) only uses the IMU measurement. Est. 1 is  used to estimate the state by~\eqref{eq x estim} for the detection as in~\eqref{est d_k}. When GPS is free of attacks, Est. 1 is also used to estimate the state for the control since the GPS measurement is trustful. In the presence of the GPS attack, Est. 2 is used for the control.}
\label{fig:est details} \vspace{-.6cm}
\end{figure}

\begin{remark}
As shown in Fig.~\ref{fig:est details}, the resilient state estimation uses the GPS measurement and the IMU measurement to estimate the state by~\eqref{eq x estim} for the detection purpose as in~\eqref{est d_k}.
When the GPS attack is detected, only the IMU measurement is used to estimate the state for the control purpose as in~\eqref{eq x estim} and~\eqref{eq P_update} with $K^G_k =0$.
\end{remark} \vspace{-.4cm}

\subsection{Attacker Location Estimation (ALT)} \label{sec: ALT}

We formulate the simultaneous estimation of the attacker location $x_k^a$ and unknown parameter $\eta_k$ as a target tracking problem of the attacker state $\vect{x}_k^a := [(x_k^a)^\top, \eta_{k}]^\top$.

Estimating the attacker state $\vect{x}_k^a$ encounters two major problems: $i)$ the output equation $y_k^S$ in~\eqref{eq1 d} is highly nonlinear, and $ii)$ a single measurement of the signal strength suffers from the infinite number of solutions.

To address the first issue, we use the unscented Kalman filter (UKF)~\cite{julier1997new,wan2000unscented}, which has been developed to deal with highly nonlinear systems and provides a better estimation than the extended Kalman filter.
Motivated by the fact that locating the epicenter of an earthquake can be done with at least three measurements from different seismic stations, we resolve the second issue by using sliding window outputs.
To be specific, we estimate $\vect{x}_{k+1}^a$ using UKF with $M$-sized sliding window outputs:
\begin{align*}
    \vect{x}_{k+1}^a = \vect{x}_{k}^a + \vect{w}_k^a; \quad
    \boldsymbol{y}_k^S  = [y_{k}^S, y_{k-1}^S, \cdots,y_{k-M+1}^S]^\top. 
\end{align*}
The signal strength measurements from~\eqref{eq1 d} can be written as $y_k^S = f(\vect{x}_{k}^a) + v_k^S$, where $f(\vect{x}_{k}^a) :=  C^S \frac{\eta_k}{d(x^a_k, x_k)^2}$.

The state estimation by using UKF with sliding window outputs can track the location of the moving attacker, while nonlinear regression algorithms may fail to track it.

For completeness of the paper, the UKF with sliding window outputs algorithm is summarized in Algorithm~\ref{algo1} in Appendix with a brief derivation.

\subsection{Escape Controller (ESC)}

In the presence of the GPS spoofing attack, the variance $P_k$ in~\eqref{eq P_update} of the state estimation errors is strictly increasing and unbounded in time (Thm. 4.2~\cite{yoon2019towards}).
The goal of ESC is to drive the UAV outside of the effective range of the spoofing device within the escape time so that the state estimation error remains within the tolerable region with a predetermined probability. The escape time is defined in~\cite{yoon2019towards} as the following:
\begin{definition}
The escape time $k^{esc} \geq 0$ is the time difference between the time of attack $k^a$ and the first time instance when the estimation error $\|x_k-\hat{x}_k\|$ is within the tolerable error distance $\zeta \in {\mathbb R}^n_+$ with the significance $\alpha$, i.e.
\begin{align}
    &k^{esc}=\arg \min_{k \geq k^a} k - k^a\nnum\\
    &\text{s.t.  } \zeta^\top P_{k}^{-1}\zeta < \chi_{df}^2(\alpha), \label{escapetime}
\end{align}
where $P_{k}$ is the error covariance of $x_k-\hat{x}_k$, $df$ is the degree of freedom of the state.
\label{def1}
\end{definition}

The escape time provides a new criterion for optimal trajectory regeneration with increasing uncertainties.
In particular, ESC is designed to drive the UAV outside of the effective range of the spoofing device within the escape time.

Given the estimates of UAV state $\hat{x}_k$ and attacker state $\hat{\vect{x}}_k^a$ with their covariances, the problem can be formulated as a finite horizon constrained MPC problem:
\begin{program}\label{mpc01}
\begin{align}
    \min_{u} \ &\sum_{i=k^a}^{k^a+N}\hat{\tilde{x}}_{i+1}^\top Q_i \hat{\tilde{x}}_{i+1}+u_i^\top R_i u_i\nonumber \\
    \text{s.t.  } &\hat{x}_{i+1} = A \hat{x}_i + B u_i \nonumber\\ %\label{eq2 mpc01 b}\\
    &d(\hat{x}^a_{k^a+k^{esc}}, \hat{x}_{k^a+k^{esc}})-{r_{\textit{effect}}}>0 \label{eq2 mpc01 c}\\
    &h(\hat{x}_i,u_i) \leq 0\label{eq2 mpc01 d}\\
    &\text{   for } i = k^a, k^a+1, \cdots, k^a+N, \nonumber
\end{align}
\end{program}
where $N \geq k^{esc}$ is the prediction horizon,
$\hat{\tilde{x}}_i$ is defined as the difference between the state estimation and the goal state at time index $i$, i.e., $\hat{\tilde{x}}_i := \hat{x}_i - x_i^{goal}$,
$Q_i$ and $R_i$ are symmetric positive definite weight matrices,
and $\hat{x}^a_i$ is the estimate of the attacker location.
Value $r_{\textit{effect}}$ is the upper bound of the effective range of the spoofing device.
The constraint~\eqref{eq2 mpc01 c} implies that ESC should drive the UAV outside of the effective range of the spoofing device.
Inequality~\eqref{eq2 mpc01 d} is any nonlinear constraint on the state estimation $\hat{x}_i$ (e.g., velocity) and the control input $u_i$ (e.g., acceleration).

\begin{remark} 
The upper bound of the effective range $r_{\textit{effect}}$ can be assumed to be known.
Due to hardware constraints, the output power/nominal strength of the spoofing device $\eta_k$ is bounded, and $\eta_k$ also can be estimated by ALT in Section~\ref{sec: ALT}.
The output power determines the effective range of the spoofing device, and $r_{\textit{effect}}$ can be found by
$r_{\textit{effect}} = \argmax_{r}  g(r)$, where $g(r): = C^S \frac{\eta_k}{r^2} > \eta^S$.
\end{remark}

There are two key challenges in Program~\ref{mpc01}.
First, the states and the attacker location are unknown and their estimates $\hat{x}_{i}$ and $\hat{x}^a_{i}$ are subject to stochastic noise.
Moreover, we cannot guarantee that constraint~\eqref{eq2 mpc01 c} is always feasible;
i.e., Program~\ref{mpc01} may not have a solution.
Addressing the above two challenges, we introduce two programs for ESC in Section~\ref{sec:detESC} and~\ref{sec:ptESC}.

\subsubsection{ESC with Tube}\label{sec:detESC}

Since the constraint~\eqref{eq2 mpc01 c} is the safety critical constraint, we can reformulate it as a conservative constraint such that ESC should drive the UAV outside of the effective range of the spoofing device with probability $\gamma$ by the single individual chance constraint (ICC):
\begin{align}
    \mathbb{P}[d(x^a_{k^a+k^{esc}}), x_{k^a+k^{esc}}]-{r_{\textit{effect}}}>0)>\gamma. \label{eq:chance constrain}
\end{align}
Then Program~\ref{mpc01} becomes a new stochastic MPC with ICC.

The chance constraints can be handled by constraint backoffs, which originate in linear MPC with additive stochastic noise~\cite{van2002conic}.
However, we consider nonlinear constraints in Program~\ref{mpc01}, which makes the backoff intractable to compute.
In~\cite{kohler2019computationally}, the tube is constructed based on  sublevel sets of the incremental Lyapunov function by an online predicted tube size and then it is used to ensure robust constraint satisfaction by tightening the nonlinear state and input constraints.
In~\cite{schluter2019constraint}, this is extended to allow for ICCs and stochastic uncertainty.
Similar to~\cite{kohler2019computationally,schluter2019constraint}, the stochastic MPC with ICC can be formulated as:
\begin{program}\label{mpc03} 
\begin{align}
    \min_{u} \ &\sum_{i=k^a}^{k^a+N}\hat{\tilde{x}}_{i+1}^\top Q_i \hat{\tilde{x}}_{i+1}+u_i^\top R_i u_i\nonumber\\
    \text{s.t.  } &\hat{x}_{i+1} = A \hat{x}_i + B u_i \nonumber \\%\label{smpcdd-cons1}\\
    &d(\hat{x}^a_{k^a+k^{esc}}, \hat{x}_{k^a+k^{esc}})-{r_{\textit{effect}}}>s(P_{k^{a}+k^{esc}},P_{k}^a,\gamma) \label{eq2 mpc dete}\\%&d(\hat{x}^a_{k^a+k^{esc}}, \hat{x}_{k^a+k^{esc}})-{r_{\textit{effect}}}>0 \\
    &h(\hat{x}_i,u_i) \leq 0 \label{smpcdd-cons3}\\
    &\text{for \ } i = k^a, k^a+1, \cdots, k^a+N, \nonumber
\end{align}
\end{program}
where $P_{k^{a}+k^{esc}}$ is the UAV state covariance at escape time, and $P_{k}^a$ is the attacker state covariance.
Function $s(\cdot)$ is the probabilistic tube size that can be seen as a margin to fulfill the second constraint in~\eqref{eq2 mpc01 c}.

In order to provide the theoretical guarantees on the capability of Program~\ref{mpc03} and the equivalence between the stochasic MPC with ICC and Program~\ref{mpc03}, we use the results from~\cite{kohler2019computationally,schluter2019constraint}.
Since the newly formulated MPC with ICC~\eqref{eq:chance constrain} is the standard nonlinear stochastic MPC problem, Assumptions in~\cite{schluter2019constraint} can be verified.

\begin{theorem}\label{thm1}
Under the Assumptions 1-4, 6 and 9 in~\cite{schluter2019constraint}, if Program~\ref{mpc03} is feasible at $t = k^a$, then it is recursively feasible; the constraints~\eqref{eq2 mpc01 d} and \eqref{eq:chance constrain} are satisfied and the origin is practically asymptotically stable for the resulting closed loop system.
The impact of the hard constraint~\eqref{eq2 mpc dete} is equivalent to the nonlinear ICCs~\eqref{eq:chance constrain}.
\end{theorem}
\begin{pf}
See proofs of Thm. 1 in~\cite{kohler2019computationally} and Thm. 8 \& 10 in~\cite{schluter2019constraint}.
\end{pf}

From Theorem \ref{thm1}, we can conclude that as long as Program~\ref{mpc03} is feasible at the time of attack $k^a$, we can guarantee that the UAV can escape within the escape time in probability.
However, Program~\ref{mpc03} may not be feasible in some cases.
To address this issue, we introduce a program with a soft constraint in the subsequent section.

\subsubsection{ESC with Potential Function}\label{sec:ptESC}

The hard constraint~\eqref{eq2 mpc dete} can be replaced by the repulsive potential function~\cite{ge2000new} as a high penalty in the cost function which is active only after the escape time $k^a + k^{esc}$.
The repulsive potential function $U_{rep}(D)$ is defined as the following:
\begin{align*}
    U_{rep}(D) :=\left\{\begin{array}{ll}{\frac{1}{2}\beta\left(\frac{1}{D}-\frac{1}{r_{\textit{effect}}}\right)^{2}} & {\text { if } D \leq r_{\textit{effect}}} \\
    {0} & {\text { if }  D>r_{\textit{effect}}}\end{array}\right.,
\end{align*}
which can be constructed based on the distance between the location of the attacker and the location of UAV, $D := d(\hat{x}_{k^a+k^{esc}}^a, \hat{x}_{k^a+k^{esc}})$.
The scaling parameter $\beta$ is a large constant, which represents a penalty when the constraint has not been fulfilled.
Utilizing the soft constraint, we reformulate the MPC problem as  follows:
\begin{program}\label{mpc04}
\begin{align*}
    \min_{u} \ &\sum_{i=k^a}^{k^a+N}\hat{\tilde{x}}_{i+1}^\top Q_i \hat{\tilde{x}}_{i+1}+u_i^\top R_i u_i + \sum_{i=k^a+k^{esc}}^{k^a+N}U_{rep}(D_i) \nonumber\\
    \text{s.t.  } &\hat{x}_{i+1} = A \hat{x}_i + B u_i \\%\label{pfinMPC}\\
    &h(\hat{x}_i,u_i) \leq 0\nonumber\\
    &\text{for \ } i = k^a, k^a+1, \cdots, k^a+N. \nonumber
\end{align*}
\end{program}
% Program~\ref{mpc04} is a non-convex nonlinear programming problem, which can be solved by the nonlinear programming algorithms such as sequential quadratic programming (SQP)~\cite{boggs1995sequential,lawrence2001computationally}.

\begin{remark} 
Comparing to the use of the repulsive potential function $U_{rep}$ in the collision avoidance literature~\cite{olfati2006flocking,choset2005principles,wolf2008artificial}, the proposed application of the repulsive potential function in Program~\ref{mpc04} has two differences.
First of all, the repulsive potential function is known before the collision happens in collision avoidance literature, while we can only get the repulsive potential function $U_{rep}$ after the collision happens, i.e., only after the UAV has entered the effective range of the spoofing device.
Second, the repulsive potential function $U_{rep}$ is only counted in the cost function in Program~\ref{mpc04} after the escape time.
\end{remark}

%%%%%%%%%%%%%%%%%%%%%%%%%%%%%%%%%%%%%%%%%%%%%%%
%%%%%%%%%%%%%%   Simulation  %%%%%%%%%%%%%%%%%%
%%%%%%%%%%%%%%%%%%%%%%%%%%%%%%%%%%%%%%%%%%%%%%%
%%%%%%%%%%%%%%%%%%%%%%%%%%%%%%%%%%%%%%%%%%%%%%%
%\clearpage
\section{Simulation} \label{sec:simulation}

In the simulations, the UAV is moving from the start position with the coordinates at $(0, 0)$ to the target position $(300, 300)$ by using feedback control\footnote{We implemented a proportional-derivative (PD) like tracking controller, which is widely used for double integrator systems.},
based on the state estimate from~\eqref{eq x estim}.
When the GPS attack happens, the state estimate will be no longer trustful.
After GPS measurement is turned off,  the only available relative state measurement causes the sensor drift problem~\cite{yoon2019towards}.
The UAV will switch the control mode from the robust control mode to the emergency control mode when the attack is detected, using ESC to escape away from the attacker within the escape time.
We solve the problem with \emph{ESC with Potential Function} described in Program~\ref{mpc04}.
The online computation is done using \texttt{Julia}, and ESC is implemented by using \texttt{JuMP}~\cite{DunningHuchetteLubin2017} package  with \texttt{Ipopt} solver.

\subsection{UAV Model}
We use a double integrator UAV dynamics under the GPS spoofing attack as in~\cite{kerns2014unmanned}.
The discrete time state vector $x_k$ considers planar position and velocity at time step $k$, i.e.
$x_k = [r_k^x, r_k^y, v_k^x, v_k^y]^\top$,
% \begin{equation*}
%     x_k = [r_k^x, r_k^y, v_k^x, v_k^y]^\top,
% \end{equation*}
where $r_k^x, r_k^y$ denote $x$, $y$ position coordinates, and $v_k^x, v_k^y$ denote velocity coordinates.
We consider the acceleration of UAV as the control input $u_k = [u_k^x, u_k^y]^\top$.
We assume that the state constraint and control input constraint are given as 
$\sqrt{(v_k^x)^2 + (v_k^y)^2}  \leq 5$ and $\sqrt{(u_k^x)^2 + (u_k^y)^2}  \leq 2$.
With sampling time at $0.1$ seconds, the double integrator model is discretized into the following matrices:
\begin{equation*}
    A = 
    \begin{bmatrix} 
    1 & 0 & 0.1 & 0    \\
    0 & 1 & 0    & 0.1 \\
    0 & 0 & 1    & 0 \\
    0 & 0 & 0    & 1
    \end{bmatrix},
    \quad
    B = 
    \begin{bmatrix} 
    0    & 0 \\
    0    & 0 \\
    0.1 & 0 \\
    0    & 0.1 
    \end{bmatrix},
\end{equation*}
and the outputs $y^G_k$, $y^I_k$ and $y^S_k$ are the position measurements from GPS, the velocity measurements from IMU, and GPS signal strength measurements respectively, with the output matrices:
$C^G = 
    \begin{bmatrix} 
    1 & 0 & 0 & 0    \\
    0 & 1 & 0 & 0
    \end{bmatrix}$,
$C^I = 
    \begin{bmatrix} 
    0 & 0 & 1 & 0 \\
    0 & 0 & 0 & 1 
    \end{bmatrix}$ and 
$C^S = 
    \begin{bmatrix} 
    1 
    \end{bmatrix}$.
The covariance matrices of the sensing and disturbance noises are chosen as
$\Sigma_w = 0.1I$, $\Sigma_G = I$, $\Sigma_I = 0.01I$ and $\Sigma_S = I$.

\subsection{GPS Spoofing Attack and Attack Signal Estimation}
\begin{figure}[thpb]
\centering
 \includegraphics[width=0.40\textwidth]{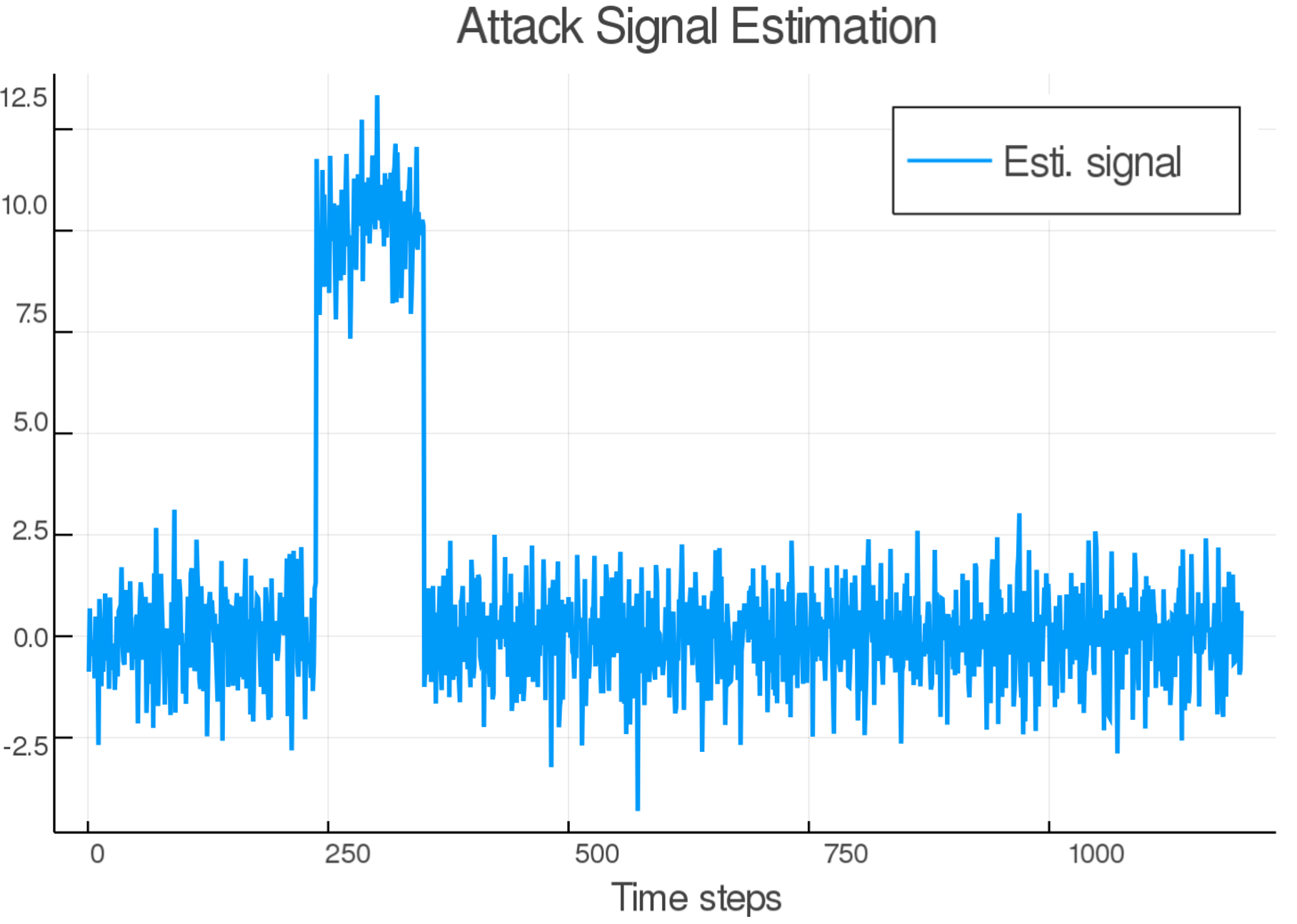}
\caption{\it Attack signal estimation. The UAV stays in the effective range of the spoofing device from time step $231$ to $356$.}
\label{fig:attack esti}\vspace{-.5cm}
\end{figure}

The GPS attack happens when the UAV is in the effective range of the spoofing device.
In this attack scenario, the attack signal is $d = [10, 10]^\top$.
The location of the attacker and the nominal power of the spoofing device are $x^a_k = [100,100]^\top$ and $\eta_k = 200$, which are both unknown to the UAV.
The estimation obtained by~\eqref{est d_k} is shown in Fig.~\ref{fig:attack esti}. 

\subsection{Attack Detection}
\begin{figure}[thpb]
\centering
 \includegraphics[width=0.40\textwidth]{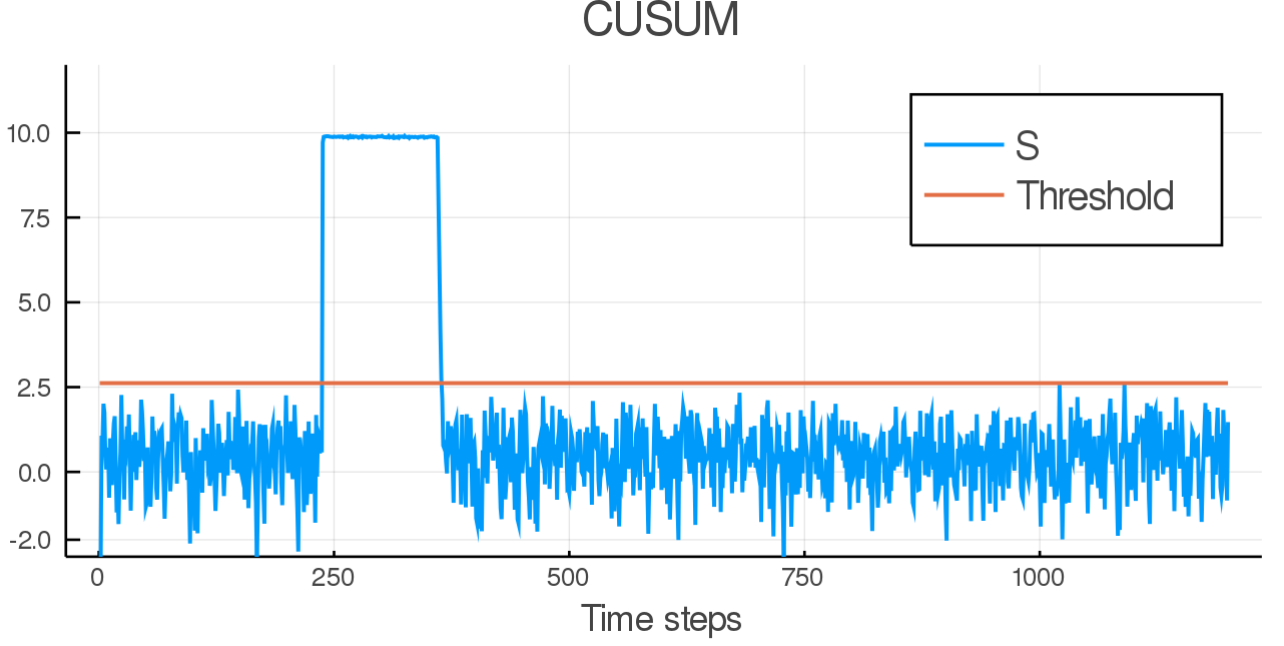}
\caption{\it Attack detection. The detector state $S_k$ is defined in~\eqref{e003.1} of the CUSUM detector. The threshold is calculated by $\frac{\chi^2_{df}(\alpha)}{1-\delta}$ with $\alpha = 0.01$ and $\delta = 0.15$.}
\label{fig: CUSUM}\vspace{-.5cm}
\end{figure}

Using the estimated attack signal to calculate the detector state $S_k$ by~\eqref{e003.1}, the
 attack detector is able to detect the attack using the normalized attack vector as shown in Fig.~\ref{fig: CUSUM}.
In Fig.~\ref{fig: CUSUM}, there are abnormal high detector state values, which imply that there is an attack.
Statistic significance of the attack is tested using the CUSUM detector described in~\eqref{e003.2} with the significance $\alpha$ at $1\%$.

\subsection{Attacker State Estimation}

\begin{figure}[thpb]
\centering
 \includegraphics[width=.9\linewidth]{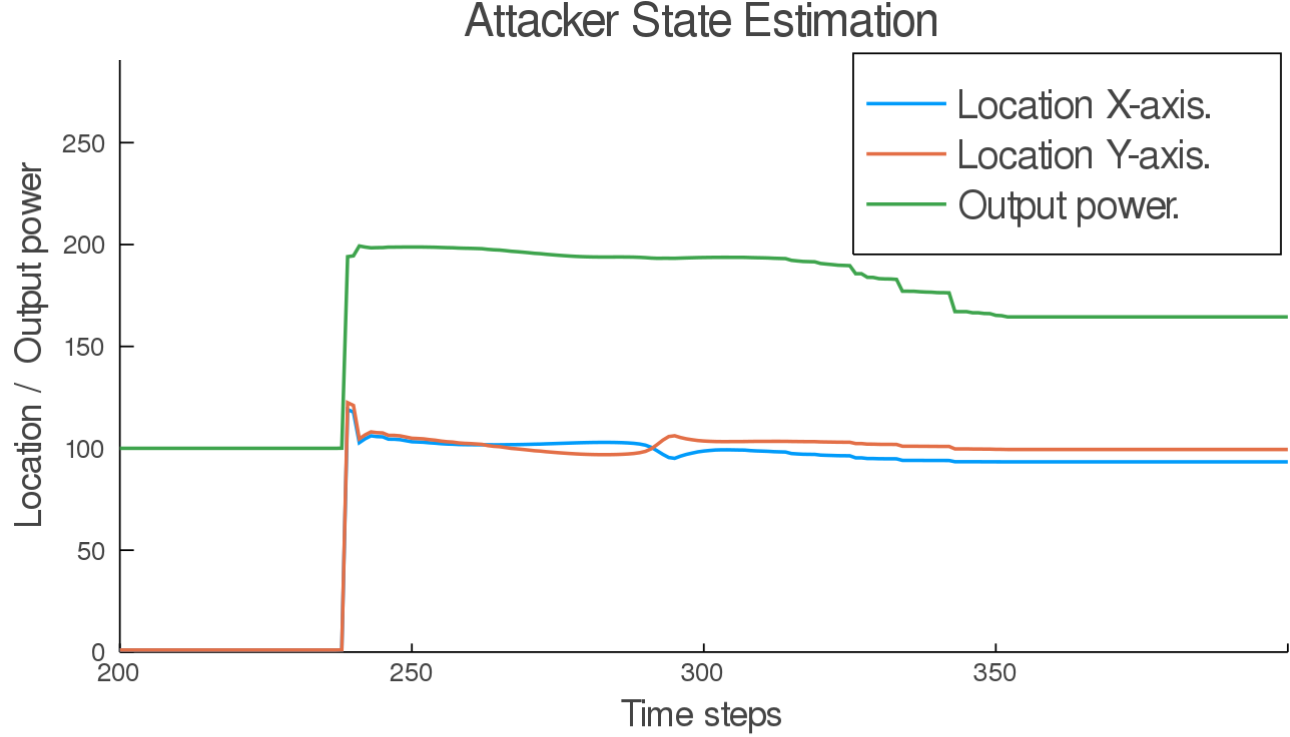}
\caption{\it Attacker state estimation.}
\label{fig: xa esti}\vspace{-.5cm}
\end{figure}

When the GPS attack is detected, the UAV first estimates the attacker state $\vect{x}_k^a$ by using Algorithm~\ref{algo1} with window size $M = 5$.
The estimation result is shown in Fig.~\ref{fig: xa esti}.
The estimated location and the estimated nominal power quickly converge to the true values.
The estimates are drifting when the UAV remains in GPS denied environment.
After obtaining an estimate of the attacker state, ESC is used to escape away from the effective range of the spoofing device.

\subsection{Trajectory Generation}
\begin{figure}[thpb]
\centering
 \includegraphics[width=0.4\textwidth]{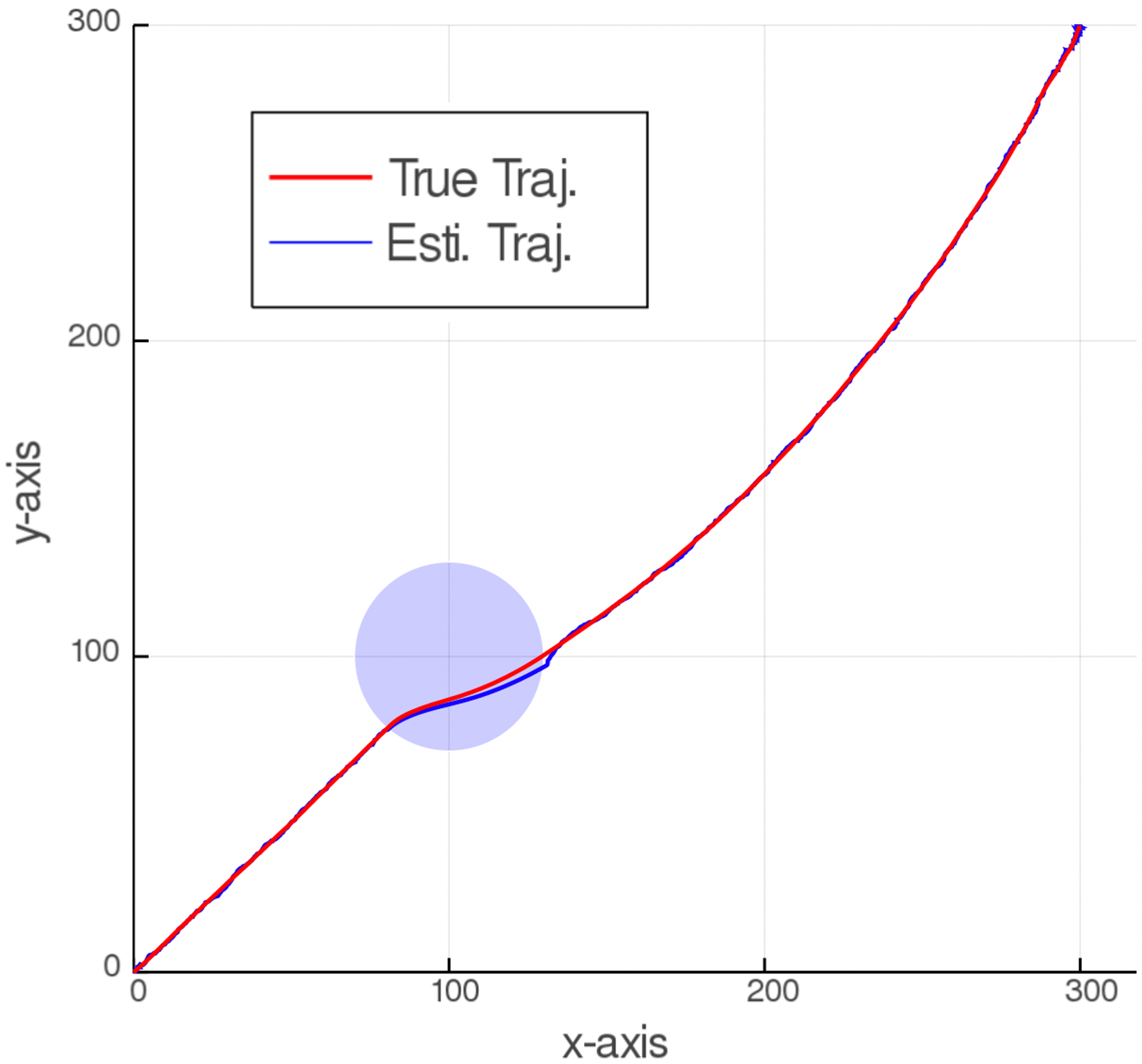}
\caption{\it Estimated and true trajectories of the simulated scenario. The attacker is located at $(100, 100)$ with $r_{\textit{effect}} = 30$, which is displayed as the light blue circle.}
\label{fig:traj}\vspace{-.4cm}
\end{figure}

\begin{figure}[thpb]
\centering
 \includegraphics[width=0.40\textwidth]{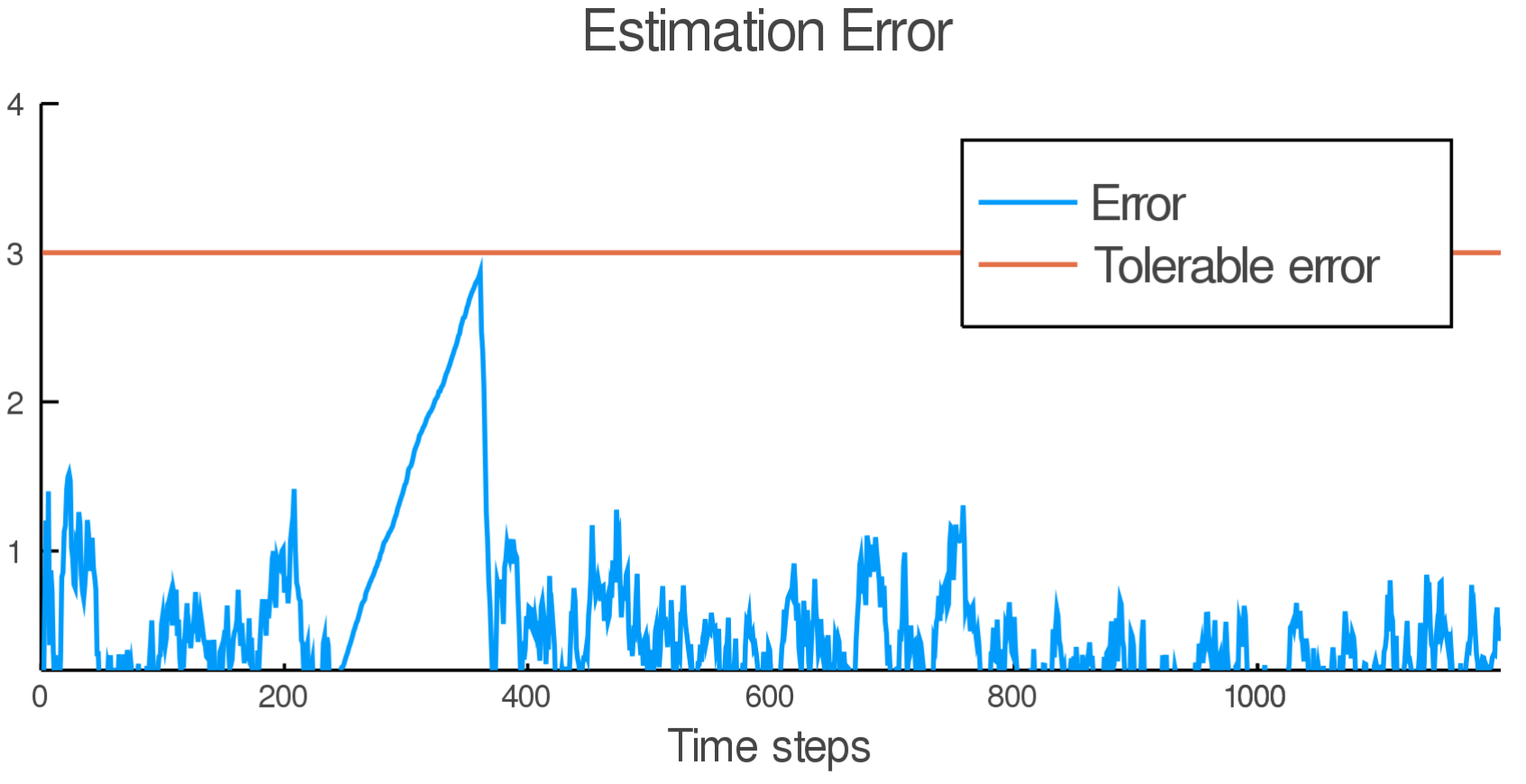}
\caption{\it Bounded estimation error $\|x_k - \hat{x}_k\|$.}
\label{fig:state error} \vspace{-.5cm}
\end{figure}

 Program~\ref{mpc04} with the prediction horizon $N = k^{esc} + 40$ and the scaling parameter $\beta = 50000$ is used to generate the estimated and true trajectories of the simulated scenario shown in Fig.~\ref{fig:traj}.
As shown in Fig.~\ref{fig:state error}, the state estimation error $\|x_k - \hat{x}_k\|$ is increasing when the UAV is in the effective range of the spoofing device, and the error is bounded by the tolerable error distance $\zeta = 3$ corresponding to $k^{esc} = 125$.

\begin{figure*}[thpb]
        \centering
        \begin{subfigure}[b]{0.2\textwidth}
            \centering
            \includegraphics[width=\textwidth]{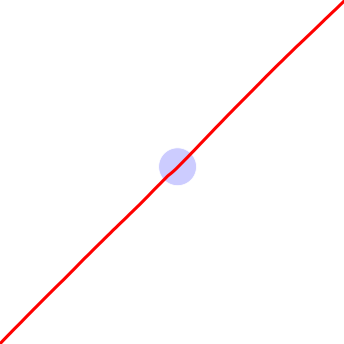}
            \caption[]%
            {{\small $r_{\textit{effect}} = 10$}}    
            \label{fig:mean and std of net14}
        \end{subfigure}
        \hfill
        \begin{subfigure}[b]{0.2\textwidth}  
            \centering 
            \includegraphics[width=\textwidth]{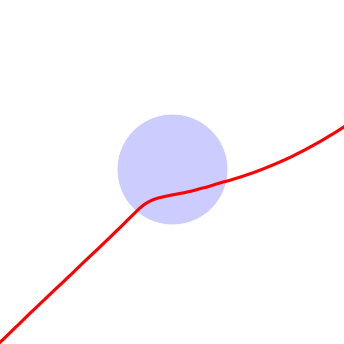}
            \caption[]%
            {{\small $r_{\textit{effect}} = 30$}}    
            \label{fig:mean and std of net24}
        \end{subfigure}
        \hfill
        \begin{subfigure}[b]{0.22\textwidth}   
            \centering 
            \includegraphics[width=\textwidth]{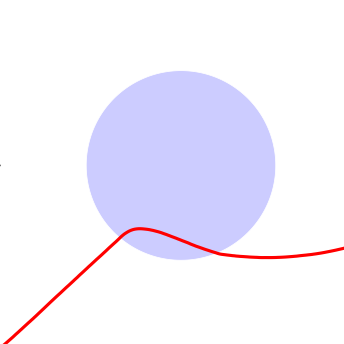}
            \caption[]%
            {{\small $r_{\textit{effect}} = 50$}}    
            \label{fig:mean and std of net34}
        \end{subfigure}
        \quad
        \begin{subfigure}[b]{0.25\textwidth}   
            \centering 
            \includegraphics[width=\textwidth]{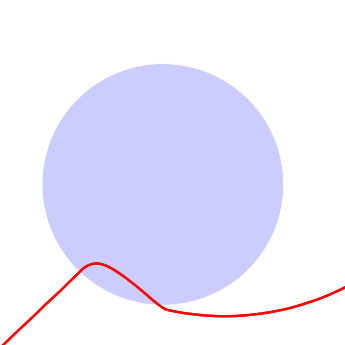}
            \caption[]%
            {{\small $r_{\textit{effect}} = 70$}}    
            \label{fig:mean and std of net44}
        \end{subfigure}
        \caption
        {\small \it Trajectories with different effective ranges. In (a), the UAV can pass the attacker without changing the direction or even its speed since $r_{\textit{effect}}$ is small enough. From (b)-(d), the UAV drives away from the effective range within the escape time and tries to get as close to the global goal as possible.  } 
        \label{fig:mean and std of nets} \vspace{-.7cm}
    \end{figure*}

Fig.~\ref{fig:mean and std of nets} presents how the proposed control framework performs in different cases where $r_{\textit{effect}} \in \{10,30,50,70\}$.
Regardless of the size of $r_{\textit{effect}}$, the UAV will escape the effective range within the escape time.

\subsection{Discussion} 

\begin{figure}[ht] 
        \centering
         \begin{subfigure}[b]{.5\textwidth}   
            \centering 
            \includegraphics[width=0.7\textwidth]{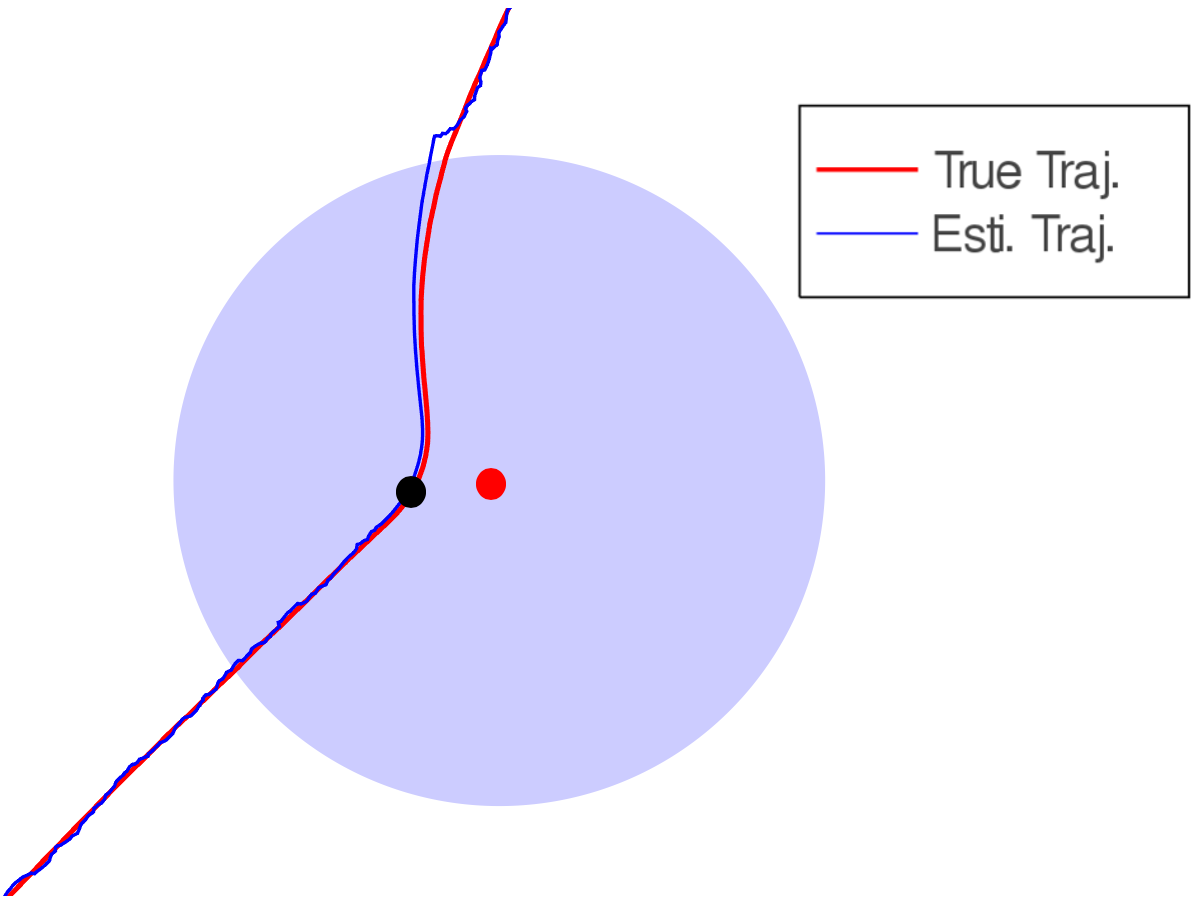}
            \caption[]%
            {{\small \it True and estimated trajectories. The spoofing device with $r_{\textit{effect}} =40$ is located at the red dot and starts to spoof when the UAV is at the black dot.}}    
            \label{fig: nonfeasiTraj}
        \end{subfigure}
        \\
        \begin{subfigure}[b]{.5\textwidth}   
            \centering 
            \includegraphics[width=.8\textwidth]{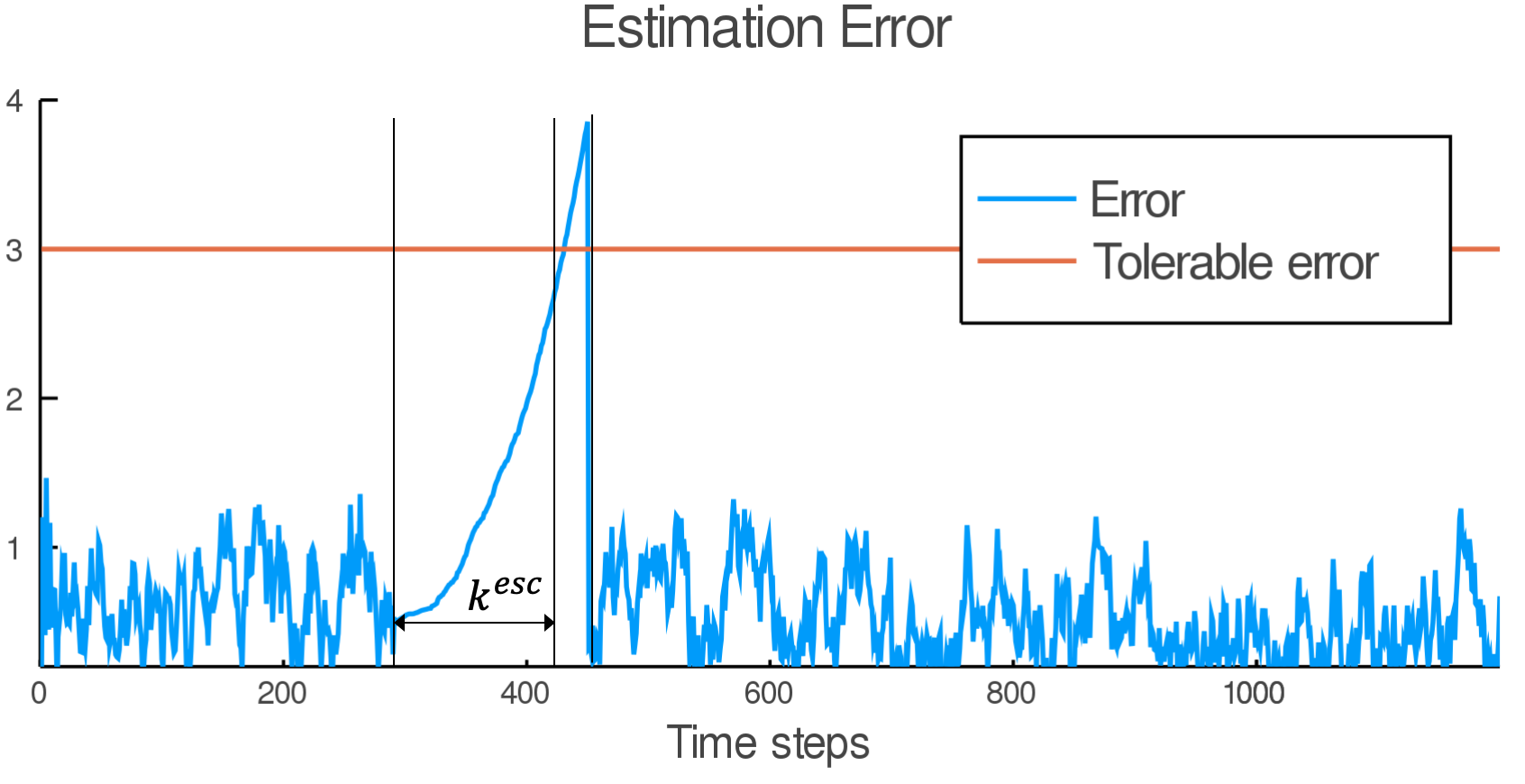}
            \caption[]%
            {{ \it Estimation error for non-feasible case. The error is bounded within the escape time. The UAV could escape at $11$ time steps after escape time at time step 423. }}    
            \label{fig: nonfeasiError}
        \end{subfigure}
        \caption
        {\it Non-feasible Case. Attacker starts the spoofing device when the UAV is nearby.} 
        \label{fig: nonfeasiCase}\vspace{-.6cm}
    \end{figure}

When the UAV is under the GPS spoofing attack, the feasibility of Program~\ref{mpc03} depends on how far the UAV is away from the boundary of the effective range of the spoofing device. However, Program~\ref{mpc04} can still generate a solution such that the UAV can escape with a minimum time, even when Program~\ref{mpc03} is not feasible.
To simulate this scenario, we now consider an attack strategy that the attacker starts the attack when the UAV is near to the spoofing device.
The results are shown in Fig.~\ref{fig: nonfeasiCase}, where Program~\ref{mpc04} generates an optimal control sequence, such that the UAV could escape at $11$ time steps after the escape time.

\section{Conclusion}

We present a secure safety constrained control framework that adapts the UAV at a path re-planning level to support resilient state estimation against GPS spoofing attacks.
A resilient state estimator has been designed, and the $\chi^2$ CUSUM algorithm is used for attack detection.
In the presence of the GPS spoofing attack, the state estimation suffers from increasing variance due to the limited sensor availability.
In this case, using the robust controller may still keep the UAV within the effective range of the spoofing device after the estimation errors may not be in the tolerable region.
The large estimation error will give rise to  safety problems. 
To solve this safety problem,
ALT is developed for tracking the attacker location and estimating the effective range of the spoofing device by using UKF with sliding window outputs.
Then, ESC is used to escape away from the effective range of the spoofing device within the escape time.
A UAV simulation is given to demonstrate the results.

%%%%%%%%%%%%%%%%%%%%%%%%%%%%%%%%%%%%%%%%%%%%%%%
%%%%%%%%%%%%.     Reference.     %%%%%%%%%%%%%%
%%%%%%%%%%%%%%%%%%%%%%%%%%%%%%%%%%%%%%%%%%%%%%%
\bibliography{ref}{}
\bibliographystyle{ieeetr}

%%%%%%%%%%%%%%%%%%%%%%%%%%%%%%%%%%%%%%%%%%%%%%%
%%%%%%%%%%%%%%%%%%%%%%%%%%%%%%%%%%%%%%%%%%%%%%%
%%%%%%%%%%%%%%%%%%%%%%%%%%%%%%%%%%%%%%%%%%%%%%%
%%%%%%%%%%%%%%%%%%%%%%%%%%%%%%%%%%%%%%%%%%%%%%%
%%%%%%%%%%%%%%%%%%%%%%%%%%%%%%%%%%%%%%%%%%%%%%%
%%%%%%%%%%%%%%%%%%%%%%%%%%%%%%%%%%%%%%%%%%%%%%%
\appendix 
The current appendix presents the UKF algorithm statement and its brief derivation for partially nonlinear systems with sliding window outputs. 
Consider the system:
% \begin{align}
%     x_{k+1}= A_kx_{k} + w_k; \quad
%     y_k  =   f(x_k) + v_k,
%     \label{eq sukf sys}
% \end{align}
\begin{align}
\begin{split}
    x_{k+1}&= A_kx_{k} + w_k\\
    y_k & =   f(x_k) + v_k,
    \label{eq sukf sys}
\end{split}
\end{align}
where $x_k \in {\mathbb R}^n$ is the state, the
output $y_k \in {\mathbb R}^{m}$ represents the measurement of GPS signal strength.
The noise signals $w_k$ and $v_k$ are assumed to be independent and identically distributed Gaussian random variables with zero means and  covariances
${\mathbb E[w_k w_k^\top]=\Sigma}_{w'} \geq 0$
and ${\mathbb E[v_k v_k^\top]=\Sigma}_v >0$.

%=====================================================
%====.     algorithm  starts      ====================
%=====================================================
\begin{algorithm}[thpb] 
\caption{UKF with sliding window outputs}\label{algo1}
\begin{algorithmic}[1]
%=====================================================
\Statex     $\rhd$ Prediction 
%=====================================================
\State $\hat{x}_{k|k-1} = A_{k-1}\hat{x}_{k-1}$;
%=====================================================
\State $P_{k|k-1} = A_{k-1}P_{k-1}A_{k-1}^\top + \Sigma_{w'}$;
%=====================================================
% \Statex
\Statex     $\rhd$ Sigma points generation
\State $\mathcal{X}_k = \{ \hat{x}_{k|k-1} \pm (\sqrt{nP_{k|k-1}})_i^\top \}$, $i\in \{1,\cdots, n\}$;
%=====================================================
\Statex      $\rhd$ Measurement Update
%=====================================================
\For{$i = 1:2n$}
%=====================================================
\State \vspace{-.7cm}
\begin{align*}
\boldsymbol{\hat{y}}^i_k &:=[\hat{y}_k^i,\hat{y}_{k-1}^i,\cdots,\hat{y}_{k-M+1}^i]^\top\\
&=[f(\mathcal{X}_k^i),(A_{k-1}^{-1}\mathcal{X}_k^i), \cdots,f(A_{k-1}^{-M+1}\mathcal{X}_k^i)]^\top;
\end{align*}
\vspace{-.7cm}
%=====================================================
\EndFor
%=====================================================
\State $\boldsymbol{\bar{y}}_k = \sum_{i=0}^{2n}W_k^i \boldsymbol{\hat{y}}_k^i$;
\vspace{+1mm}
%=====================================================
\State $\boldsymbol{P}_k^{y}=\sum_{i=0}^{2n}W_k^i(\boldsymbol{\hat{y}}_k^i-\boldsymbol{\bar{y}}_k)(\boldsymbol{\hat{y}}_k^i-\boldsymbol{\bar{y}}_k)^\top + {\boldsymbol{\Sigma}_{v}}$;
\vspace{+1mm}
%=====================================================
\State $\boldsymbol{P}_k^{xy} = \sum_{i=0}^{2n}W_k^i (\mathcal{X}_k^i-\hat{x}_{k|k-1})(\boldsymbol{\hat{y}}_k^i-\boldsymbol{\bar{y}}_k)^\top$;
\vspace{+1mm}
%=====================================================
\State $K_k = \boldsymbol{P}_k^{xy}(\boldsymbol{P}^y_k)^{-1}$
\vspace{+1mm}
%=====================================================
\State $\hat{x}_k=\hat{x}_{k|k-1}+K_k(\boldsymbol{y}_k-\boldsymbol{\bar{y}}_k)$;

%=====================================================
\State $P_k = P_{k|k-1} - K_k \boldsymbol{P}_k^y K_k^\top  $
%=====================================================
\end{algorithmic}
\end{algorithm}
%=====================================================
%========.     algorithm  ends     ================
%=====================================================
\vspace{-.4cm}

\subsection{Algorithm Derivation} \label{sec append}

\noindent \emph{Prediction.}
Given the previous state estimate $\hat{x}_{k-1}$ and system model~\eqref{eq sukf sys}, the current state can be predicted as $\hat{x}_{k|k-1} = A_{k-1}\hat{x}_{k-1}$. 
% \begin{align*}
%     \hat{x}_{k|k-1} = A_{k-1}\hat{x}_{k-1}.
% \end{align*}
Its error covariance matrix is
$P_{k|k-1} := \mathbb{E}[(x_{k}-\hat{x}_{k|k-1})(x_{k}-\hat{x}_{k|k-1})^\top]= A_{k-1}P_{k-1}A_{k-1}^\top + \Sigma_{w'}$,
% \begin{align*}
%     P_{k|k-1} :&= \mathbb{E}[(x_{k}-\hat{x}_{k|k-1})(x_{k}-\hat{x}_{k|k-1})^\top]\\
%     &= A_{k-1}P_{k-1}A_{k-1}^\top + \Sigma_{w'},
% \end{align*}
where $P_{k-1}:= \mathbb{E}[(x_{k-1}-\hat{x}_{k-1})(x_{k-1}-\hat{x}_{k-1})^\top]$ is the state estimation error covariance matrix. 

\noindent \emph{Sigma Points Generation.}
We define a sigma points array
$\mathcal{X}_k := \{ \hat{x}_{k|k-1} \pm (\sqrt{nP_{k|k-1}})_i^\top , i=1,\cdots, n\}$,
% \begin{align*}
%     \mathcal{X}_k := \{ \hat{x}_{k|k-1} \pm (\sqrt{nP_{k|k-1}})_i^\top , i=1,\cdots, n\} ,
% \end{align*}
where $\sqrt{nP_{k|k-1}}$ is the matrix square root such that $\sqrt{nP_{k|k-1}}^ \top\sqrt{nP_{k|k-1}} =nP_{k|k-1}$, and the matrix operator $(\cdot)_i$ gives the $i^{th}$ row of the matrix.

\noindent \emph{Measurement Update.}
Given the sliding window size $M$, the nonlinear measurement equation $f(\cdot)$ is used to transform the sigma points into predicted measurement vectors
$\boldsymbol{\hat{y}}^i_k :=[\hat{y}_k^i,\hat{y}_{k-1}^i,\cdots,\hat{y}_{k-M+1}^i]^\top=[f(\mathcal{X}_k^i),f(A_{k-1}^{-1}\mathcal{X}_k^i),\cdots,f(A_{k-1}^{-M+1}\mathcal{X}_k^i)]^\top$.

%\begin{align*} \hat{y}_k^i &= f(\mathcal{X}_k^i)\\ \hat{y}_{k-1}^i &= f(A_{k-1}^{-1}\mathcal{X}_k^i)\\ &\vdots\\ \hat{y}_{k-N+1}^i &= f(A_{k-1}^{-N+1}\mathcal{X}_k^i), \end{align*}

% {\color{green}
% The approximated mean of the measurements is defined by
% {\blue$
%     \boldsymbol{\bar{y}}_k := \sum_{i=0}^{2n}W_k^i \boldsymbol{\hat{y}}_k^i,
% $}
% where $W_k^i$ are the weighting coefficients.}

The approximated mean of the measurements is 
$
    \boldsymbol{\bar{y}}_k := \sum_{i=0}^{2n}W_k^i \boldsymbol{\hat{y}}_k^i,
$
where $W_k^i$ are the weighting coefficients.

By taking the measurement noise into account, the estimated covariance of the predicted measurements is given by:
$
    \boldsymbol{P}_k^{y}:=\sum_{i=0}^{2n}W_k^i(\boldsymbol{\hat{y}}_k^i-\boldsymbol{\bar{y}}_k)(\boldsymbol{\hat{y}}_k^i-\boldsymbol{\bar{y}}_k)^\top + {\boldsymbol{\Sigma}_{v}},
$
where ${\boldsymbol{\Sigma}_{v}} = \diag\{ \Sigma_{v},\cdots,  \Sigma_{v} \}$ is the diagonal matrix.

The cross covariance between the state prediction and predicted measurements is 
$
    \boldsymbol{P}_k^{xy} = \sum_{i=0}^{2n}W_k^i (\mathcal{X}_k^i-\hat{x}_{k|k-1})(\boldsymbol{\hat{y}}_k^i-\boldsymbol{\bar{y}}_k)^\top ,
$
where $\mathcal{X}_k^i$ denotes the $i^{th}$ element in $\mathcal{X}_k$.

The measurement $\boldsymbol{y}_k:=[y_k,\cdots,y_{k-M+1}]^\top$ is used to update the prediction $\hat{x}_{k|k-1}$ as $\hat{x}_k=\hat{x}_{k|k-1}+K_k(\boldsymbol{y}_k-\boldsymbol{\bar{y}}_k)$.

The covariance matrix of the state estimation error is $
    P_k = P_{k|k-1} - K_k(\boldsymbol{P}_k^{xy})^\top - \boldsymbol{P}_k^{xy}K_k^\top + K_k \boldsymbol{P}_k^{y} K_k^\top.
$
The gain matrix $K_k$ is chosen by minimizing the trace norm of $P_k$; i.e. $\min_{K_k}\trace{(P_k)}$. The solution of the program is given by
$K_k = \boldsymbol{P}_k^{xy}(\boldsymbol{P}^y_k)^{-1}.
$
Note that the prediction step does not need unscented transformation because the dynamic system~\eqref{eq sukf sys} is linear.

\end{document}